\documentclass[journal]{IEEEtran}

\usepackage{graphicx}%
\usepackage{dcolumn}%
\usepackage{bm}%

\usepackage[utf8]{inputenc}
\usepackage[T1]{fontenc}
\usepackage{textcomp}
\usepackage{mathptmx}
\usepackage{etoolbox}
\usepackage{orcidlink}

\begin{document}
    \title{Frequency-selective membrane-based metamaterial filters for low-insertion-loss harmonic crosstalk suppression in sub-harmonic THz receivers}
  \author {Hui~Yuan \orcidlink{0000-0002-1677-6837},
      Meng~Zhang, Almila~Kura, 
      Daniel~Erni \orcidlink{0000-0002-1467-6373}
      and~Hartmut~G.~Roskos \orcidlink{0000-0003-3980-0964}

  \thanks{This work has been submitted to the IEEE for possible publication. Copyright may be transferred without notice, after which this version may no longer be accessible.}
  \thanks 
  {H. Yuan and H. G. Roskos are with the Physikalisches Institut, Goethe-Universität Frankfurt am Main, Frankfurt am Main, Germany (e-mail:yuan@physik.uni-frankfurt.de; roskos@physik.uni-frankfurt.de).}
  \thanks{M. Zhang and D. Erni are with General and Theoretical Electrical Engineering (ATE), CENIDE – Center for Nanointegration Duisburg–Essen, University of Duisburg-Essen, Duisburg, Germany}
  \thanks{A. Kura is with Department of Physics, Bilkent University, Ankara, Turkey.}}  

\markboth{}{Yuan \MakeLowercase{\textit{et al.}}: Frequency-selective membrane-based metamaterial filters}

\maketitle

\begin{abstract}
Parasitic reflections and standing waves fundamentally limit amplitude stability and phase fidelity in coherent continuous-wave (CW) terahertz (THz) systems. We report a complementary metamaterial filter pair designed for frequency-selective isolation at 300~GHz and 600~GHz in sub-harmonic heterodyne architectures. Fabricated on ultra-thin polyimide membranes, the filters provide more than 30~dB stopband isolation while maintaining insertion losses below 1~dB in the passband.
System-level coherent-imaging experiments demonstrate effective suppression of spurious-reflection-induced interference effects and improved spectral baseline stability. The filters mitigate wavefront distortions resulting from such interferences and enhance reconstruction robustness. 
These results establish membrane-based metamaterial filters as compact frequency-selective isolation elements for 
sub-harmonic heterodyne THz detection 
and highlight their potential for phase-engineered imaging and sensing applications.
\end{abstract}

\maketitle

\section{Introduction}

Coherent %
CW %
THz systems have become increasingly important for spectroscopy and imaging due to their capability to simultaneously measure amplitude and phase of the electromagnetic field \cite{989974, Ferguson2002,glaab_terahertz_2010,s21124092}. %
The phase sensitivity enables precise extraction of dielectric properties and high-resolution Fourier-domain imaging \cite{yuan_3d_2019, Yuan:23, 10.1063/1.4819734}. However, coherent operation also makes such systems inherently vulnerable to parasitic reflections and standing-wave formation. Multiple quasi-optical interfaces within the RF and local oscillator (LO) beam paths create coupled interference channels that perturb both amplitude stability and phase fidelity, ultimately degrading spectral baselines and reconstruction accuracy in imaging applications \cite{10.1063/5.0217898, proceedings2025129051}.

Frequency-selective surfaces (FSSs) and metamaterial-based filters have emerged as powerful tools for manipulating THz wave propagation \cite{cai2010optical, wong2020activea, Jiang:17, Ghosh:22}. Recent reports have demonstrated flexible \cite{WU2023115563}, tunable \cite{wrro170228, app10134660, 9706347, Zhang:25, VASIC2024115377}, and polarization-sensitive \cite{ PANG2023128975, Li:25} metasurfaces capable of amplitude modulation, phase control, and spectral filtering in the sub-THz and THz regimes. The metal-mesh/inductive-capacitive filtering concept itself is well established \cite{Ade2006metalmesh}; the novelty here lies in its system-level application -- embedding such filters within a coherent sub-harmonic receiver for harmonic crosstalk suppression (cf.\ Appendix~\ref{appendixc}).

In coherent sub-harmonic detection architectures,  %
harmonic crosstalk between RF and LO channels constitutes a particularly critical challenge. Because both frequency components propagate within the same quasi-optical environment, residual leakage and reflection can lead to strong coherent coupling between channels. Traditional absorptive filtering approaches introduce insertion and reflection loss and may compromise phase coherence in power-limited THz systems. A reflective, low-loss isolation strategy that preserves phase stability while selectively suppressing unwanted harmonics is therefore highly desirable.

The present work employs polyimide (PI) as the supporting substrate for the metamaterial structures. PI, having a rather high melting point (above 600~K), is favored for flexible printed circuits, semiconductor insulation, and the fabrication of insulating tapes due to its mechanical strength and %
structural stability. In the THz range, PI exhibits very low opacity and minimal dispersion over a broad spectral bandwidth \cite{9566999, LIU2019109264}. With an absorption constant of approximately 2 cm$^{-1}$ at 300 GHz to 7 cm$^{-1}$at 600 GHz, a 5-$\mu$m-thick membrane yields an absorption loss of only 
0.005–0.015 dB across the operational frequency band. The real part of the refractive index ranges between 1.8 and 1.9 over this frequency range, \cite{Sahin2019}. %
and the sub-wavelength 
membrane ($d/\lambda < 1\%$) suppresses the net reflection loss to below 0.03 dB. The combined intrinsic loss therefore remains well below 0.05 dB across the operation band, negligible compared to the sub-1-dB insertion-loss budget of the filters. This combination of physical, electrical, mechanical, and optical properties makes PI an ideal candidate material for the design and fabrication of flexible metamaterial THz filters.

In this work, we present a complementary pair of flexible metamaterial-based frequency-selective filters designed for operation at 300~GHz and 600~GHz in coherent sub-harmonic heterodyne THz platforms. Fabricated on ultra-thin PI membranes, the filters operate predominantly in a reflective stopband regime, redirecting unwanted frequency components out of the optical path rather than absorbing them. We demonstrate that this approach not only achieves strong harmonic isolation with minimal insertion loss, but also modifies the interference topology of the receiver architecture in a geometry-dependent manner. These findings establish membrane-based metamaterial filters as compact and mechanically flexible elements for reflection management and phase stabilization in advanced coherent THz systems.

\section{Metamaterial Filter Design and System-Level Implementation}
\label{Meta}

As discussed in the previous section, suppressing parasitic reflections and standing waves in coherent THz systems requires isolation strategies that preserve phase coherence while introducing minimal insertion and reflection loss. To address this requirement, we employ a complementary metamaterial-based FSS  concept that operates predominantly through reflective stopband behavior rather than absorptive attenuation. By redirecting unwanted frequency components out of the optical path, the proposed approach enables efficient harmonic isolation in power-limited THz systems.

The specific application scenario at hand is that of a Fourier imaging system employing a single, raster-scanned sub-harmonic receiver for the coherent detection of 600-GHz waves \cite{Yuan:23, yuan2022_IWMTS}. The detector is a Si CMOS TeraFET (antenna-coupled field-effect transistor \cite{ojefors, lisauskas_rational_2009,ikamas_broadband_2018}) on a weakly doped Si substrate. The transistor is embedded in a broadband bow-tie antenna, which enables the detector to cover the frequency range of interest here \cite{ikamas_broadband_2018}. Attached to the device are a superstrate and a substrate lens for the injection of the object-illuminating radiation at 600~GHz and the local-oscillator (LO) beam at 300~GHz \cite{yuan2023superstratelensa, krysl_si_2024}. As illustrated in Fig.~\ref{fig1}(a), %
the two beams impinge from opposite sides onto the receiver. In contrast to the common one-sided injection, this avoids the need for a beam-combiner which would result in a loss of half of the power of each beam (in the case of 50:50 beam-splitting ratio of the combiner). The beams are not entirely absorbed in the receiver. The purpose of the inclusion of the FSS filters is to block the remainder of both beams from propagation in free space, which can lead to back-reflection, respectively back-scattering, to the receiver facilitated by the confocal nature of the superstrate- and substrate-lens arrangement. One filter (F1 in Fig.~\ref{fig1}(b)) is placed in the optical path of the 600-GHz channel to block residual 300-GHz radiation, while F2 (in Fig.~\ref{fig1}(c)) is placed in the 300-GHz channel to suppress 600-GHz radiation. This complementary FSS filtering approach is not limited to THz imaging but is applicable for other scenarios of sub-harmonic heterodyne THz detection to suppress unwanted frequency components and interferences.

The filter design is based on a complementary pair of metamaterial structures fabricated on ultra-thin PI membranes, first reported in conference form in \cite{10299203}. Each filter in the pair is designed to suppress harmonic radiation at a different target frequency: 300~GHz for the first filter (F1) and 600~GHz for the second filter (F2). The PI membrane, with a thickness of 5~$\mu$m, provides enough mechanical stability required for integration into quasi-optical THz systems, while its minimal effective optical thickness reduces dielectric loss. The constant thickness preserves radiation phase stability. The metamaterial patterns define the frequency-selective behavior and enable high stopband isolation combined with low insertion loss in the passband.

\subsection{Complementary Filter Geometry and Implementation}
\begin{figure}
\includegraphics[width=9cm]{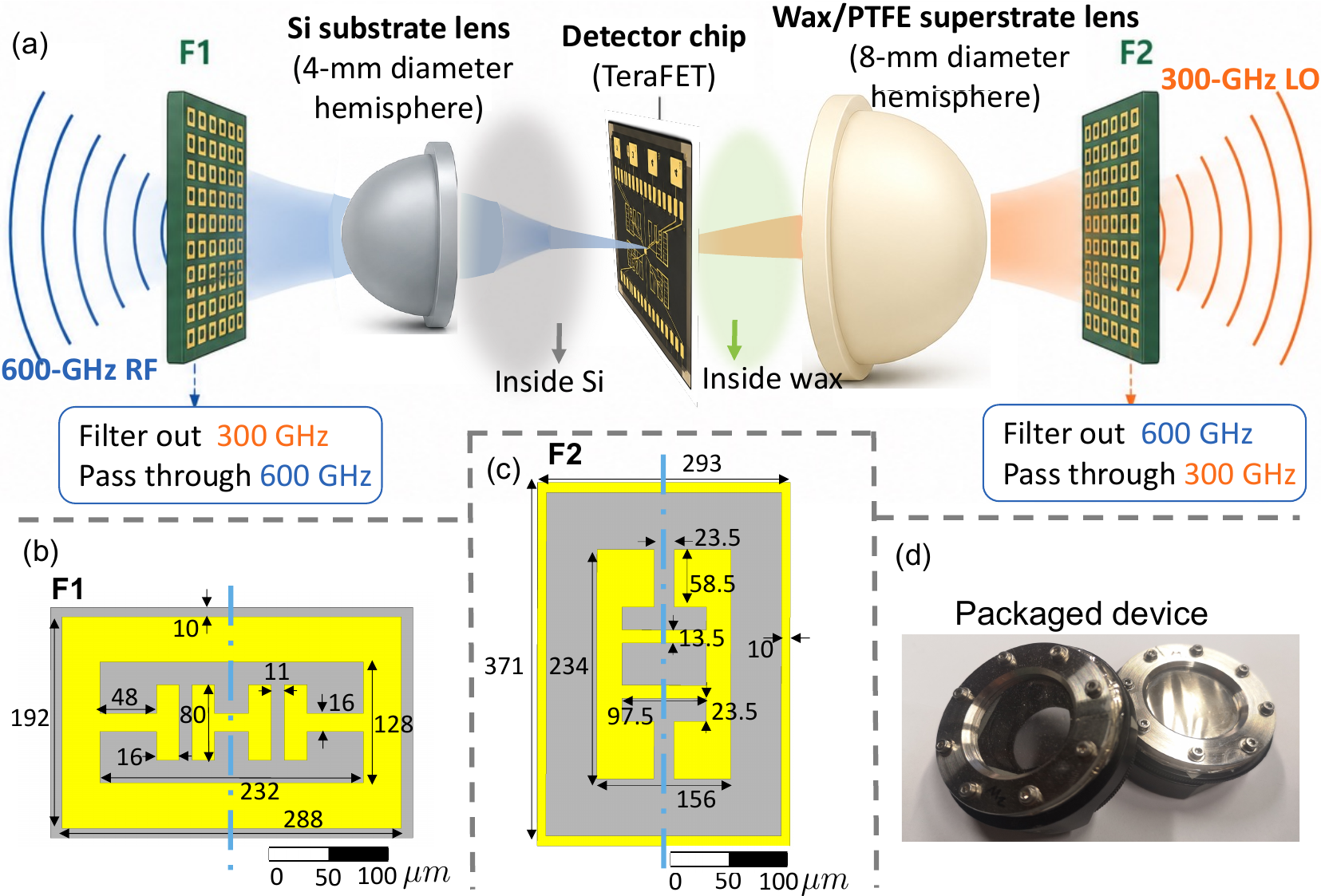}
\caption{(a): Illustration of the placement of the filters relative to the detector in a coherent sub-harmonic THz receiver configuration. (b) and (c): Design of the complementary metamaterial filters F1 and F2 with labeled geometrical dimensions and the principal axes (blue dot-dash line) respectively. (d): Photographs of the mounted filter membranes.}
\label{fig1}
\end{figure}

The left side of Fig.~\ref{fig1} illustrates the designs and geometrical dimensions of the complementary filters F1 and F2. The unit-cell dimension of F1 is 308~$\mu$m $\times$ 212~$\mu$m, while for F2, it is  371~$\mu$m $\times$ 293~$\mu$m. These dimensions were optimized to achieve strong suppression at the respective target frequencies. Filter F1 consists of two pairs of metallic bars (shaped as the circuit symbol of a capacitor) connected to a rectangular frame, forming a resonant structure that blocks radiation around 300~GHz while transmitting 600~GHz. Filter F2 is complementary to F1 in spectral response by suppressing radiation around 600~GHz while transmitting 300~GHz. We use ``complementary'' here in this functional sense rather than in the strict Babinet sense: the two unit cells are not geometric negatives of one another (F1: 308~$\mu$m~$\times$~212~$\mu$m; F2: 371~$\mu$m~$\times$~293~$\mu$m), so an exact Babinet duality is not claimed. This complementary design enables efficient harmonic isolation while maintaining low insertion loss for the desired signal band. %

The filters were fabricated directly on a 5-$\mu$m-thick PI membrane, which was deposited on a 500-$\mu$m-thick Si wafer to provide mechanical support during processing. The PI membrane was prepared by  spin-coating of a solution of ZKPI 30222 (vendor: POME) onto the Si wafer  followed by a multi-step 
thermal curing process according to the manufacturer's specifications: 
80\,\textdegree C for 2\,h, ramping to 130\,\textdegree C over 
50\,min and holding for 2\,h, then to 160\,\textdegree C over 30\,min 
(2\,h), 210\,\textdegree C over 50\,min (3\,h), 230\,\textdegree C 
over 20\,min (3\,h), and finally 270\,\textdegree C over 40\,min with 
a final hold of 10\,h.
 
A 500-nm-thick aluminum layer on top of 20-nm-thick titanium adhesion layer was subsequently deposited onto the PI membrane, and the metamaterial patterns were defined using standard photolithography and lift-off processing. After patterning, the PI membrane with the  filter structure was carefully %
pulled off from the Si carrier and mounted on a supporting steel frame. Slight, careful stretching of the membrane on the frame avoided wrinkling of the membrane. Finally, the membrane was clamped between the frame and a cover ring. %
A photograph of the packaged filters is shown on the right-hand side of Fig.~\ref{fig1}.

\subsection{Operating Principle and Numerical Simulations}

The frequency-selective behavior of the complementary metamaterial filters was investigated using full-wave electromagnetic simulations performed with CST Microwave Studio. Each filter was modeled as a periodic array of unit cells patterned on a 5~$\mu$m-thick PI membrane, using the tabulated conductivity of aluminum. Periodic boundary conditions were applied in the transverse directions and open (radiation) boundary conditions along the propagation direction. Plane-wave illumination under normal incidence was used to determine the frequency-dependent transmission and reflection characteristics.

\begin{figure}
\includegraphics[width=8 cm]{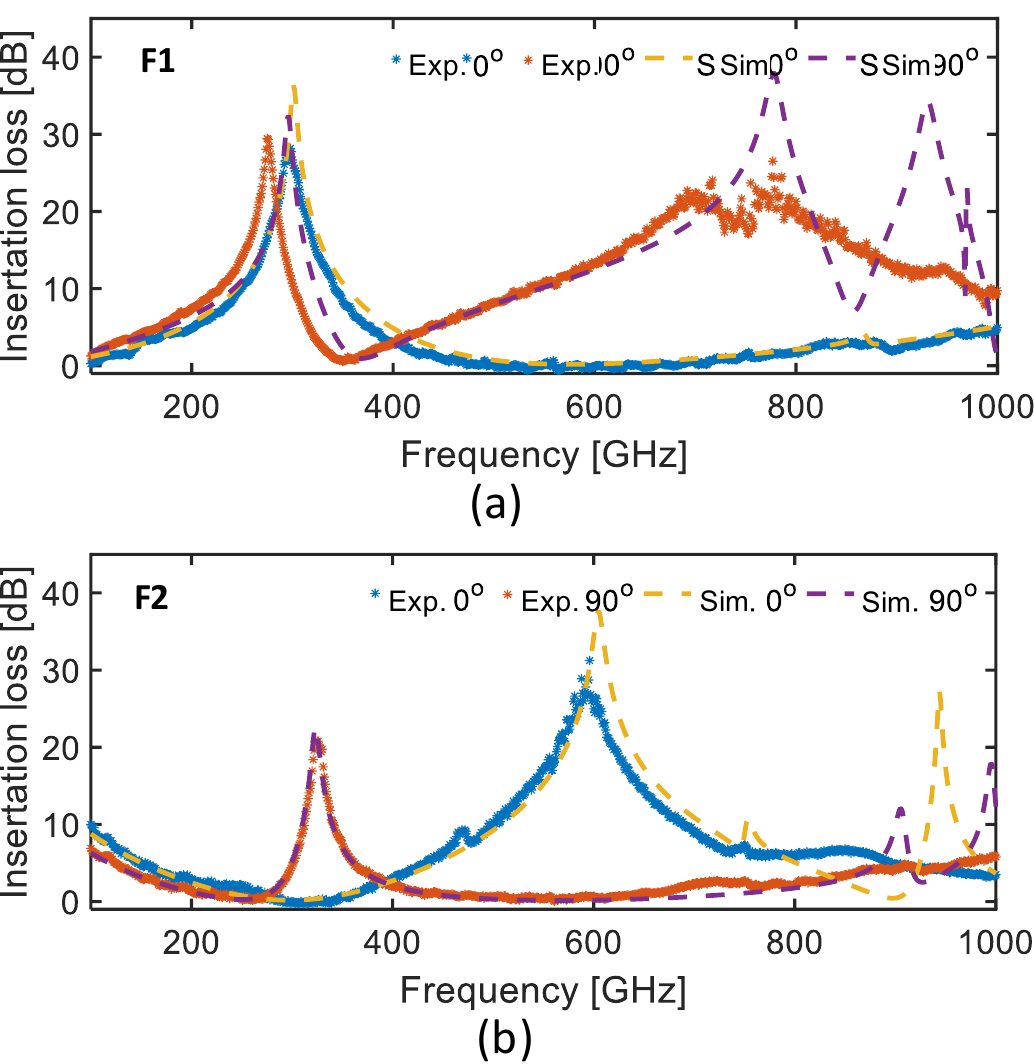}%
\caption{ 
Transmission characteristics of the complementary metamaterial filters. Data shown for polarization directions of 0$^\circ$ and 90$^\circ$ with respect to the principal axes shown in Fig.~\ref{fig1}. 
Dashed lines: simulations, dots: measured results. (\textbf{a})
Filter F1 (optimized for suppression of radiation at 300~GHz and high transmission at 600~GHz for 0$^\circ$ ), (\textbf{b}) Filter F2 (optimized for suppression of radiation at 600~GHz  and high transmission at 300~GHz for 0$^\circ$ ). %
\label{fig2}}
\end{figure}

Figure~\ref{fig2} (dashed lines) illustrates the simulated transmission characteristics of the complementary filter pair under linearly polarized excitation. Fig.~\ref{fig2}(a) and (b) show the transmission spectra of~F1 and~F2 for two orthogonal polarization orientations (0$^\circ$ and 90$^\circ$ relative to the principal axes of the metamaterial structures). %
For polarization at 0$^\circ$, each filter exhibits a pronounced transmission minimum at its respective target frequency while maintaining high transmission in the complementary passband. 
The graphs plot the insertion loss $L_T$ which is defined as the transmittance expressed in dB, $L_T = -10\,\log_{10} (T)~\mathrm{dB}$. In both cases, the simulated stopband isolation exceeds 30~dB and the insertion loss in the passband remains below 1~dB.

\begin{figure}
\includegraphics[width=8 cm]{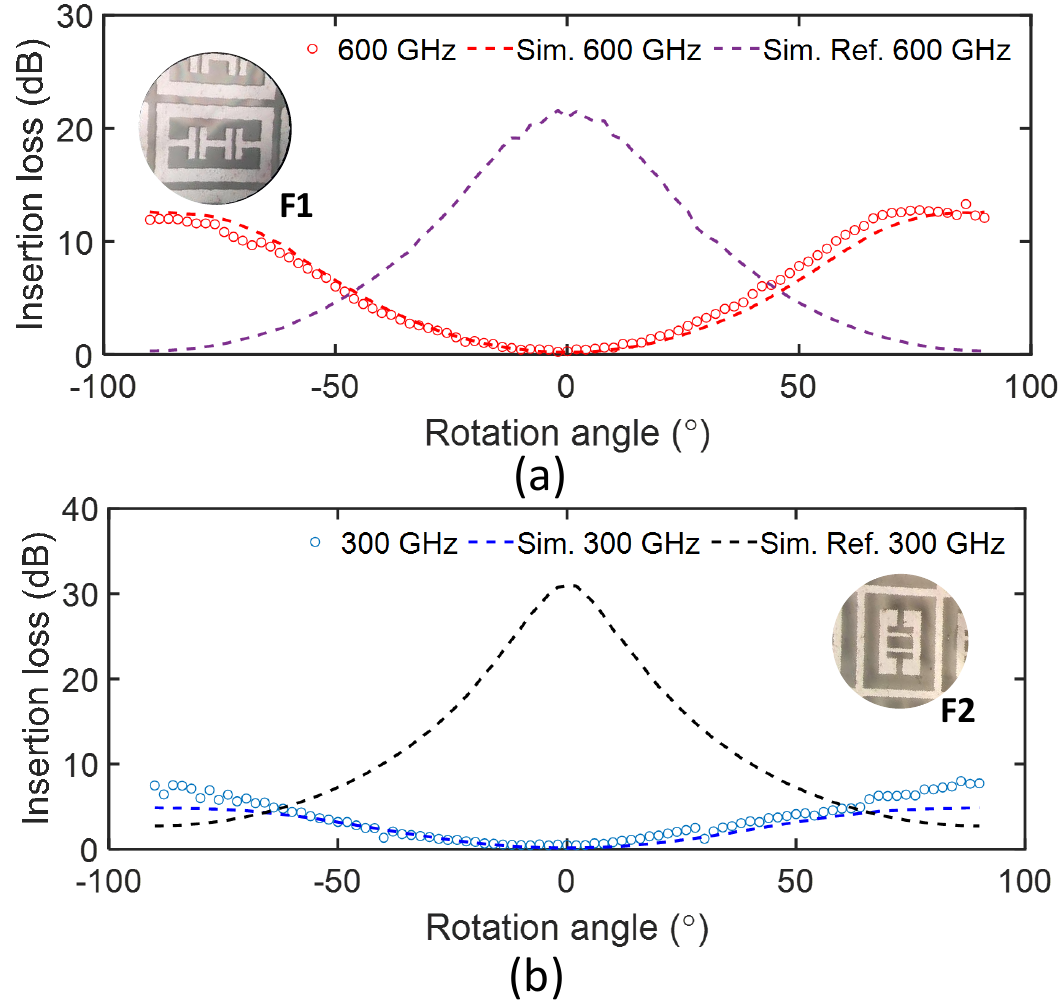}
\caption{%
Measured and simulated rotation-angle-dependent transmission characteristics of the complementary filters, (\textbf{a}) at 600~GHz for F1 (measured: dots, simulated: red dashed line, inset: microscopy photo), and (\textbf{b}) at 300 GHz for F2 (measured: dots, simulated: blue dashed line, inset: microscopy photo). In addition, simulated reflection parameters are displayed (purple, respectively black dashed line). The data correspond to the scattering parameters $S_{21}$ and $S_{11}$, respectively. 
\label{fig3}}
\end{figure}

Figure~\ref{fig3} displays simulation results of the insertion loss $L_T$ and of the reflection parameter $L_R = -10 \, \log_{10} (R)~\mathrm{dB}$ ($R$ being the reflectance) -- where the transmittance $T = |S_{21}|^2$ and the reflectance $R = |S_{11}|^2$ correspond to the squared magnitudes of the complex scattering parameters $S_{21}$ and 
$S_{11}$, respectively. %
-- as a function of the angle of the polarization of the radiation field. The simulations 
include ohmic losses in the aluminum layer and dielectric absorption
in the PI membrane; both contributions are negligibly small across
the 300--600~GHz band, so that $|S_{21}|^2 + |S_{11}|^2 \approx 1$
holds throughout. A direct consequence is that the broadband reflectance is well approximated by $R(f)\approx1-T(f)$, so the broadband transmission spectra already shown in Fig.~\ref{fig2} directly imply the corresponding broadband reflection spectra. %
The simulation data (dashed lines) are shown for the respective passband frequencies of F1 and F2, respectively. Both the transmission and reflection characteristics demonstrate that the coupling strength between the radiation and the metamaterials varies strongly with the polarization orientation. %
The polarization dependence is analogous for the two filters, albeit at the different frequencies of their passbands, 600~GHz in the case of F1 and 300~GHz for F2. This selective response enables controlled redistribution of transmitted and reflected power by adjusting the polarization orientation. The filters function as frequency-dependent beam-splitting elements, providing tunable isolation and power balancing without introducing additional absorptive losses.

The ultra-thin polyimide substrate introduces negligible Fabry-Perot resonances within the frequency range of interest. The small effective optical thickness minimizes internal multiple reflections and associated phase distortions, preserving the phase integrity of the transmitted signal. These numerical results establish the dual functionality of the complementary filter pair: Optimized bandstop isolation and polarization-tunable power redistribution.

\section{THz Characterization of the Fabricated Filter Pair}

To experimentally validate the numerical predictions and evaluate the performance of the fabricated complementary filter pair under realistic operating conditions, two-stage THz measurements were carried out. 
First, transmission measurements were performed with each filter to verify the simulated spectral responses and polarization dependencies. Second, the filters were integrated into a sub-harmonic heterodyne THz system to assess their influence on harmonic isolation and spectral stability in a coherent detection configuration.

\subsection{Frequency Response and Validation of the Polarization Sensitivity}
The fabricated filters were characterized using a CW photomixing THz system (Toptica TeraScan 1550). The measurements were performed over the frequency range from 100~GHz to 1.0~THz. Transmission spectra were recorded for both ~F1 and~F2 under linearly polarized excitation.

Figure~\ref{fig2}(a) and (b) present the measured transmission spectra (dots in Fig.~\ref{fig2}) for a polarization orientation of 0$^{\circ}$ and 90$^{\circ}$. The experimental results confirm the complementary spectral behavior predicted by the simulations. ~F1 exhibits a pronounced transmission minimum centered around 300~GHz while maintaining high transmission near 600~GHz, whereas ~F2 shows the inverse response. The measured stopband isolation exceeds 30~dB, and the insertion loss in the passband remains below 1~dB, in good agreement with the numerical results.

To further investigate the polarization-dependent behavior, the filters were rotated with respect to the incident polarization direction. The measured insertion losses are shown in Fig.~\ref{fig3}(a) and (b) as a function of the rotation angle. The measured results (dots) agree well with the simulated results (dashed lines), with some deviation between the two at large rotation angles. %

\subsection{Power Detection and Sub-Harmonic Heterodyne Detection with and without Filters in the Imaging Setup without Object}
\label{3.2} 

The primary objective of the complementary filter pair is to suppress parasitic reflections and standing waves that are particularly detrimental to phase-sensitive coherent measurements. In sub-harmonic heterodyne systems, standing-wave-induced interference not only distorts the amplitude baseline but also introduces deterministic phase distortion (reflection-induced, repeatable with frequency/position -- distinct from stochastic, temporal phase noise), which can severely impair the extraction of dielectric material properties.

Measurements were performed with the Fourier imaging set-up of Fig.~\ref{fig5}(e) described later in the main text and in the Appendix. No object-to-be-imaged was inserted and the receiver was held at a fixed position (no raster-scanning as during imaging). As stated in Sec.~\ref{Meta}, the receiver used in this work was a Si CMOS TeraFET detector. The chip was mounted on a printed circuit board. THz radiation impinged on the detector from opposite sides (cf. Fig.~\ref{fig1}), the 600-GHz RF wave through a commercial substrate lens (made from high-resistivity Si) and the 300-GHz LO wave through a home-made superstrate lens (made from wax and PTFE \cite{Yuan:23,yuan2023superstratelensa}), respectively. This dual-lens configuration enhances detection sensitivity but simultaneously opens transmission  pathways for residual radiation on both sides of the detector. Reflected waves from optical components external to the detector unit propagate back through these pathways and interfere coherently, promoting the formation of standing waves. 

The radiation sources were two frequency-tunable multiplier-chain electronic emitters with horn antennas, the LO source (vendor: Virginia Diodes Inc.) providing 600~$\mu$W of power at 300~GHz and the RF source (vendor: RPG Radiometer Physics GmbH) delivering 56~$\mu$W of power at 600~GHz. Spectral scans were performed around these center frequencies over spectral regions of approximately 100~GHz bandwidth. After placing Filter~F1 (with and without F2) into the RF  optical path, spectral analysis was conducted in the range from 235~GHz to 325~GHz. Conversely, after positioning Filter~F2 (with and without F1) in the LO optical path, measurements were made from 475~GHz to 645~GHz. The distance of the filters from the detector chip was approximately 1~cm. The filters were tilted by 30-45$^\circ$ to avoid back-reflections from them into the detector. Within this range, the exact tilt angle of each filter was set manually as a trade-off between maximizing the passband transmission (at 600~GHz for F1, at 300~GHz for F2) and keeping the specularly reflected beam clear of the optical path, rather than fixing a predetermined angle; this procedure is consistent with the angle dependence of the passband transmission discussed in Appendix~\ref{appendixb}. Full-wave simulations at the operational tilt (Appendix~\ref{appendixb}) show that the tilting leads to only minor effects on the spectral characteristics of F1 over the full $0$--$45^\circ$ range and of F2 up to $30^\circ$; however, F2's stopband shifts substantially (by $\approx$180~GHz) at $45^\circ$.

\begin{figure*}
\includegraphics[width=0.9\textwidth]{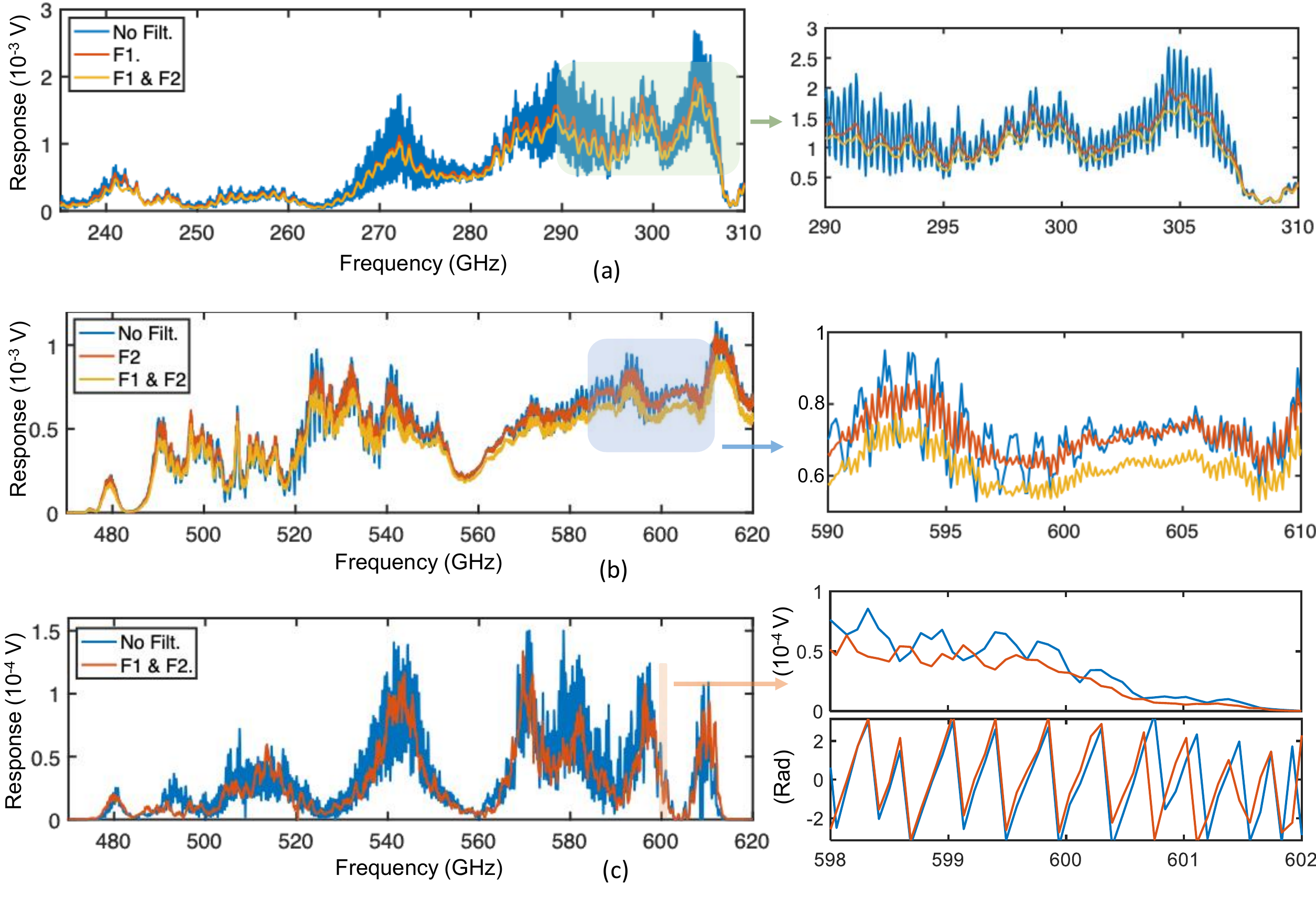}
\caption{
Spectral performance of the coherent CW THz system with and without the filters. 
(\textbf{a}, \textbf{b}) Pure power-detection spectra in the 300~GHz and 600~GHz bands acquired without filters, with a single filter, and with the complementary filter pair. 
(\textbf{c}) Complex amplitude--frequency response obtained by sub-harmonic heterodyne detection, illustrating suppression of standing-wave-induced artifacts when the filter pair is present; the lower sub-panel shows the measured phase (rad), a deterministic sawtooth ramp nearly identical with/without filters below $\approx$601~GHz but more filter-affected above, where the amplitude (upper sub-panel) weakens.\label{fig4}}
\end{figure*}

Figure~\ref{fig4}(a) and (b) present the pure power-detection spectra in the 300-GHz and 600-GHz bands, acquired respectively without filters, with a single filter, and with the complementary filter pair. Fig.~\ref{fig4}(c) shows the amplitude–versus-frequency response obtained by sub-harmonic heterodyne detection. Here, both the LO- and the RF-frequencies were swept simultaneously keeping the intermediate-frequency (IF) frequency $f_{IF} = f_{RF} - 2f_{LO} $ constant. 
In all three panels, the $y$-axis displays the as-measured voltage response from the detector resp. receiver. The measured data were not corrected for the pronounced intensity changes observed upon frequency sweeps of the emitters themselves \cite{Holstein_PhD}, the frequency-dependent changes  manifesting in the pronounced modulations of the measured voltage responses in Fig.~\ref{fig4} on a frequency scale of 5-10~GHz. Also, no correction was made for absorption of radiation by water vapor in air (see, e.g., the strong signal attenuation around 560~GHz).    

We first discuss salient results of the power-detection measurements, a more detailed discussion is found in the Appendix. In the absence of filtering, as shown by the blue lines in Fig.~\ref{fig4}(a) and (b), the resulting intensity modulation depth reaches up to 30\% for both the RF and LO waves, with the modulations being more pronounced in the RF signal for unknown reasons.

In the 300-GHz band, insertion of ~F1 significantly suppresses the standing-wave oscillations across the entire scanned bandwidth. In the spectral region directly around 300~GHz, residual oscillations are especially weak, while they are stronger at higher and lower frequencies within the 235--325~GHz scan window. This spectral behavior is attributed to the intentionally narrowband design of the filter, which is optimized to provide strong isolation at 300~GHz and 600~GHz rather than broadband attenuation: as Fig.~\ref{fig2}(a) shows directly, the stopband isolation of F1 is highest at line center and falls off away from it, so less ripple is suppressed toward the edges of the scanned window. The additional inclusion of F2 has only a marginal effect, as expected, because it is designed to let radiation in the 300-GHz region pass.

In the 600-GHz band, the introduction of ~F2 also reduces the standing-wave features, particularly near the design frequency, for the same reason (cf.\ Fig.~\ref{fig2}(b)). However, the overall suppression across the full bandwidth is slightly less pronounced than in the 300-GHz case. The additional placement of F1 does not lead to further improvement; again, this is expected because F1 is designed to transmit radiation in the 600-GHz pass.

Table~\ref{tab:ripple} quantifies this suppression using the delay-domain ripple amplitude (Fig.~\ref{figdelay}) as a common numerical metric. The dominant peak in each band is suppressed by 32.8--33.3~dB (300-GHz band, by F1) and 22.7--25.2~dB (600-GHz band, by F2) relative to the unfiltered case -- consistent with, and for the 300-GHz band slightly exceeding, the $>$30~dB stopband isolation measured at normal incidence in Fig.~\ref{fig2}; the somewhat lower suppression in the 600-GHz band is plausibly related to the reduced stopband isolation expected at the 30--45$^\circ$ operating tilt (see response to angled-incidence characterization). These suppressed peaks are the receiver-level RF$\leftrightarrow$LO crosstalk named in the title (cross-leakage mechanism, Appendix): 300-GHz (LO) into the RF-side path (F1) and 600-GHz (RF) into the LO-side path (F2). The weaker, residual peak in each band (the filter-independent self-reflection loop confined to the respective other arm, cf.\ Fig.~\ref{figa1}) remains essentially unchanged by filtering, as expected since neither filter blocks radiation in its own passband. Repeated scans were not acquired, so uncertainties on the suppression values are estimated by sub-band resampling: each single raw sweep is split into four contiguous frequency sub-bands, the ripple amplitude at the same target path length is recomputed in each, and the resulting across-sub-band spread is propagated through the suppression ratio (Table~\ref{tab:ripple}). The resulting $\pm$1.9--2.5~dB uncertainties do not change any of the qualitative conclusions drawn from the table.

\begin{table}[t]
\centering
\caption{Delay-domain ripple amplitude and suppression (dB), with and without the filters (see Sec.~\ref{3.2} for the uncertainty estimate).}
\label{tab:ripple}

\begin{tabular}{l l c c c}
\hline\hline
Band sweep & Filter & \multicolumn{2}{c}{Ripple amplitude (a.u.)} & Suppression\\
 (cf.\ Fig.~\ref{fig4}) &  & $L=13.7$~cm & $L=59.1$~cm & (dB)\\
\hline
300-GHz (a) & None  & 0.0609 & \textbf{0.1584} & -- \\
            & F1    & 0.0597 & 0.0036 & 32.8 $\pm$ 1.9 \\
            & F1+F2 & 0.0506 & 0.0034 & 33.3 $\pm$ 2.1 \\
\hline
600-GHz (b) & None  & \textbf{0.0530} & 0.0028 & -- \\
            & F2    & 0.0039 & 0.0029 & 22.7 $\pm$ 2.2 \\
            & F1+F2 & 0.0029 & 0.0027 & 25.2 $\pm$ 2.5 \\
\hline\hline
\end{tabular}
\end{table}

A more detailed discussion is presented in the Appendix. There, we address the origin of the two modulation frequencies of about 1~GHz and 250~MHz observed upon closer inspection of Fig.~\ref{fig4}(a) and (b), and the effect which the filters have on these two modulation features. The corresponding path-length assignment is independently validated there by a delay-domain (Fourier-transform) analysis of the raw spectra (Fig.~\ref{figdelay}).

We now turn to the results shown in  Fig.~\ref{fig4}(c) obtained in the sub-harmonic heterodyne detection mode. 
When both F1 and F2 are employed simultaneously, standing-wave artifacts are strongly suppressed over a broad frequency range. In particular, in the vicinity of the 600~GHz design frequency, the oscillatory modulation becomes nearly indiscernible, as highlighted in the zoomed-in view. The effect is somewhat masked by the small signal in this region, but it remains existent upon numerical magnification of the signal. This demonstrates that our frequency-selective reflection management effectively reduces deterministic amplitude and phase distortion in coherent heterodyne detection while maintaining minimal insertion loss. This reduction results from the suppression of returning spurious radiation, which is especially detrimental for phase measurements because interference between the direct and reflected waves produces a reproducible, reflection-induced phase error -- not stochastic phase noise.

At this point, we can also address the question of possible back-reflections from the filters into the detector from an experimental point-of-view. The question  arises from the reflective nature of the FSS filtering process. The concern is that a substantial part of the radiation reflected by each of the filters may return back into the detector and cause interference effects. Concluding from the spectra in Fig.~\ref{fig4}, one only sees the desired weakening of spectral modulations when the filters are placed in the beam path, and no new modulation patterns. Apparently, the divergent character of the beams leaving the detector and the positioning of the filters at a strong tilt (30-45$^{\circ}$ from the  optical axis of the measurement system) prevent back-coupling of any substantial amount of the radiation into the detector despite the proximity of the filters to the detector.

\section{Coherent CW THz Imaging}

Having demonstrated that the complementary filter pair effectively suppresses standing-wave-induced distortions of intensity and  phase information, %
we now investigate its impact on coherent CW THz imaging. %
The effectiveness of the complementary filter pair was further evaluated in a coherent CW THz Fourier imaging system  operated in sub-harmonic heterodyne mode \cite{Yuan:23}. The measurement system is shown schematically in Fig.~\ref{fig5}(e). As stated above, the object-illuminating radiation at 600~GHz and the LO radiation at 300~GHz impinged onto the receiver from opposite sides (see Fig.~\ref{fig1},  replicated in simplified form in Fig.~\ref{fig5}(d)). In the imaging system, the receiver together with the LO source (labeled "S2" in Fig.~\ref{fig5}(e)) was mounted on a translation stage which allowed to raster-scan the receiver over the focal plane of the optical system (here lens "L2" in in Fig.~\ref{fig5}(e)) to coherently acquire the Fourier-space spectrum, from which images are reconstructed by digitally applying a 2D inverse Fourier transform. As in the spectral measurements of the last chapter, standing waves and phase noise represent major limitations, particularly for high spatial-frequency components that  inherently exhibit low intensity.

\subsection{Interference Effects of the Local-Oscillator Radiation}
In heterodyne detection, the LO plays a pivotal role, as the intermediate-frequency output of the detector is directly proportional to the LO field. Consequently, spatial nonuniformity and reflections of the LO within the system can introduce severe distortions in the reconstructed images. To evaluate %
these effects, measurements were performed with the RF source switched off while the LO was active, and the LO power distribution in the focal plane of lens L2 was recorded for different filter configurations.

\begin{figure*}
\centering
\includegraphics[width=0.9\textwidth]%
{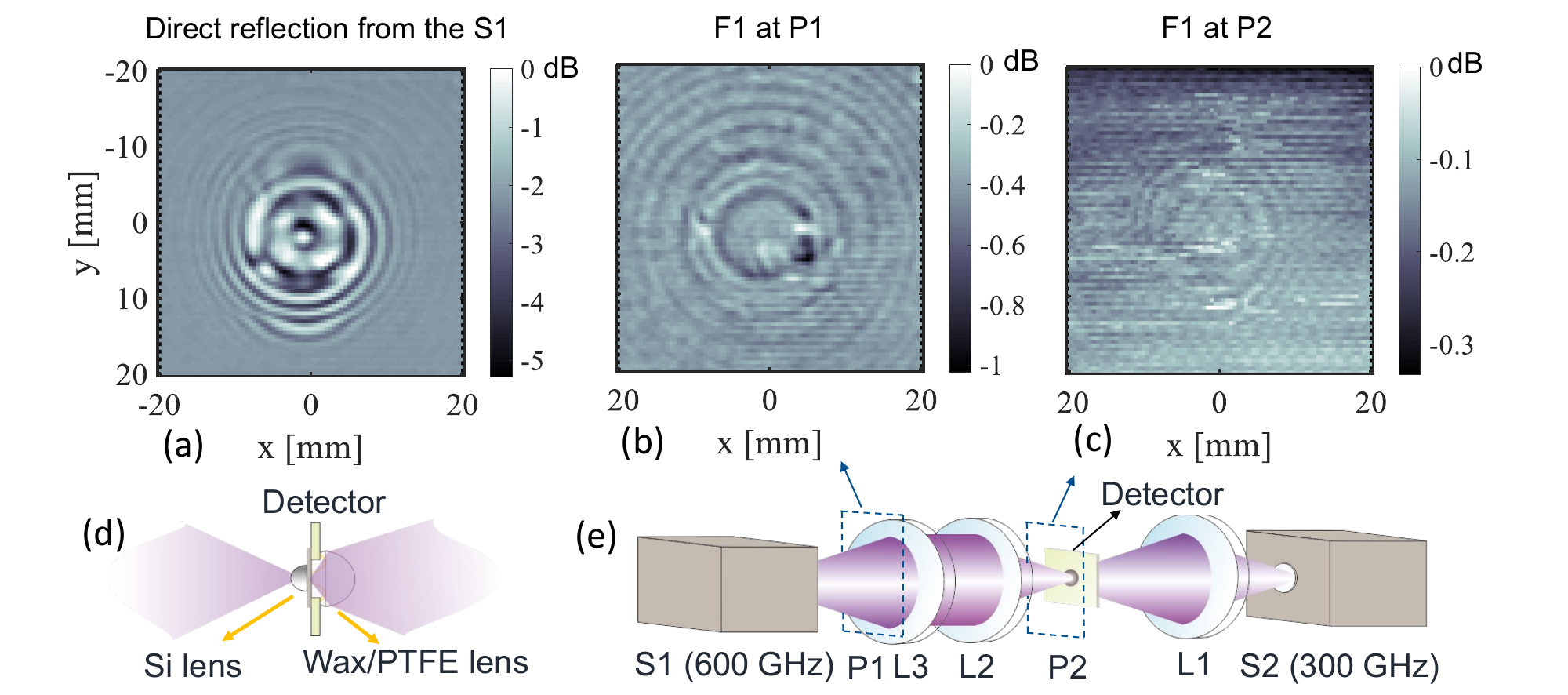}
\caption{%
Suppression of LO reflections in the sub-harmonic CW THz Fourier imaging system. 
(\textbf{a}--\textbf{c}) Measured LO power distribution in the focal plane without filtering and with F1 placed at two different positions (P1 and P2 in \textbf{e}) in the RF beam path. 
(\textbf{d}, \textbf{e}) Schematic of the coherent CW THz imaging system and the corresponding receiver configuration, indicating the positions of the complementary filters.
\label{fig5}}
\end{figure*}

Figure~\ref{fig5} shows the measured LO power distribution without filtering (Fig.~\ref{fig5}(a)) and with F1 placed at two different positions within the RF beam path: P1 (Fig.~\ref{fig5}(b)) and P2 (Fig.~\ref{fig5}(c)), the filter positions of (b,c) being displayed in Fig.~\ref{fig5}(e). In the absence of any filter, pronounced interference patterns, mostly ring-shaped and with intensity variations of approximately 5~dB, are observed, indicating strong LO reflections within the system. Positioning F1 upstream in the RF beam path at position P1 significantly reduced the intensity modulation to about 1~dB, while placing the filter directly in front of the detector at P2 suppressed the ring-shaped patterns to below 0.1~dB. Note the different grey scales of the intensity plots; the fine-grained intensity modulation seen in (c) would also be visible in (a,b) at the small modulation scale of (c).
These results demonstrate that strategically positioned frequency-selective isolation effectively eliminates dominant LO reflection pathways.

\subsection{Imaging Demonstrations}

\begin{figure*}
\centering
\includegraphics[width=0.9\textwidth]{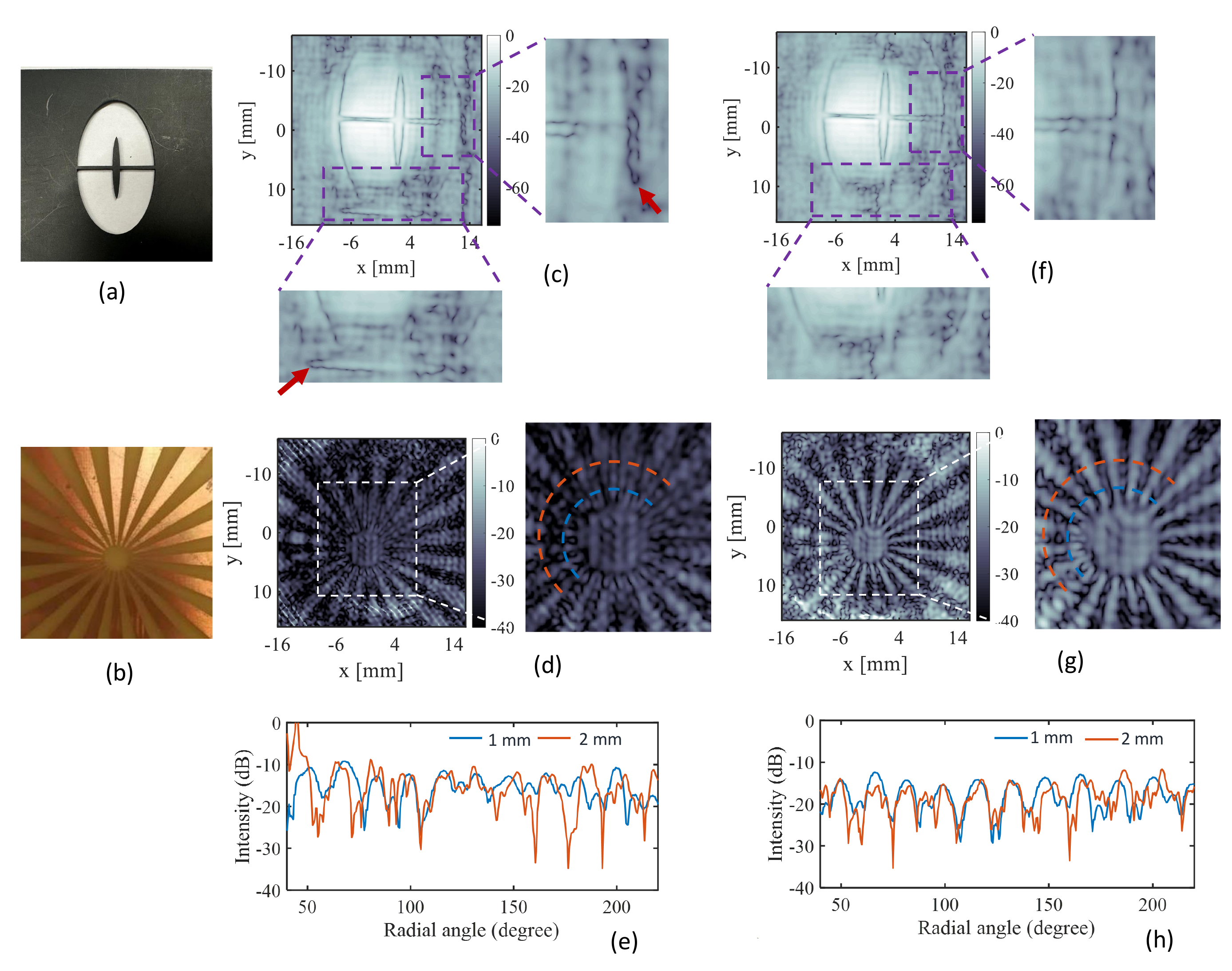}
\caption{Coherent CW THz Fourier-imaging results obtained with and without the complementary filter pair. 
(\textbf{a}, \textbf{b}) Photographs of the weakly reflective phase sample (the white color in the open region of the object is from the sheet of paper onto which has the object has been placed for the photograph) and of the highly reflective metallic Siemens star on PCB substrate, respectively. 
(\textbf{c}, \textbf{d}) Reconstructed images acquired without the filter pair, and (\textbf{f}, \textbf{g}) corresponding images acquired with the filters inserted. 
(\textbf{e}, \textbf{h}) One-dimensional tangential profiles extracted from the zoomed regions in (\textbf{d}) and (\textbf{g}). 
Profiles corresponding to metallic stripe widths of 1~mm and 2~mm are indicated by blue and red markers, respectively. %
 \label{fig6}}
\end{figure*}

To assess the impact of spurious-reflection suppression on image quality, three Fourier imaging experiments were conducted. Fig.~\ref{fig6} displays results of two of them. Two representative samples were used: a weakly reflective phase object and a highly reflective metallic object. The weakly reflective sample consisted of numerals engraved into a 1-mm-thick black plastic substrate shown in Fig.~\ref{fig6}(a), while the highly reflective sample was a metal-plated Siemens star patterned on a printed circuit board, as presented in Fig.~\ref{fig6}(b). 

In the imaging experiments, the object was illuminated by the 600-GHz beam collimated via lens L3 (cf. Fig.~\ref{fig5}(e)), and the object was placed between L3 and L2, positioned 13~cm in front of lens~L2 (focal length 6.5~cm). The dynamic range of the imaging system under these conditions was approximately 30~dB. For Fourier imaging, the detector scanned the focal plane of L2. The filters, if inserted, were placed immediately in front and behind the detector unit, with a distance of about 1~cm to it. Again, they were tilted by 30-45$^\circ$. Source power, integration time, scan step, and reconstruction parameters were kept identical with/without filters.

Figure~\ref{fig6} compares reconstructed images obtained with and without the complementary filter pair. For the weakly reflective sample, images acquired without filtering exhibit pronounced spurious artifacts and irregular shadows, to the extent that genuine structural features of the object are obscured. This effect is particularly evident at the right and bottom edges of the digit ``0'' in Fig.~\ref{fig6}(c), where the true object boundaries are masked by interference-induced artifacts. Fig.~\ref{fig6}(c) and (d) shows the full field-of-view reconstruction, with magnified insets highlighting the obscured regions further discussed below.

It is interesting to note that the artifacts are most pronounced in vignetted regions away from the optical axis. In Fourier imaging with collimated illumination, vignetting arises from two primary mechanisms. First, the illumination beam can be approximated as a quasi-Gaussian beam, whose intensity decreases toward the periphery of the beam profile, leading to a reduced signal-to-noise ratio in off-axis regions. Second, for off-center spatial Fourier frequencies, the lens may not collect the full angular spectrum symmetrically: while higher spatial frequencies are captured on one side of the Fourier space, their counterparts on the opposite side are partially lost due to the finite numerical aperture. This asymmetric loss of high spatial-frequency components further exacerbates the vignetting effect and increases sensitivity to phase noise.

As a result, deterministic phase distortion induced by spurious reflections in the beam path leads to particularly severe distortions in off-axis regions, where high spatial-frequency information is both weaker and more vulnerable to coherent interference. With the complementary filters in place, these distortions are substantially suppressed. As shown in Fig.~\ref{fig6}(f), %
spurious artifacts are effectively reduced, resulting in more physically consistent object boundaries. Notably, the bottom edge of the digit ``0'', which is far from the optical axis and therefore more strongly affected by vignetting, becomes cleaner after filtering, although the edge remains blurred due to reduced signal strength and incomplete recovery of high spatial-frequency components. However, the false interference patterns -- visible as the turbulence-like structure in the lower zoomed panel of Fig.~\ref{fig6}(c), indicated there by a red arrow -- are removed
In contrast, the right edge of the digit, which is closer to the optical axis than the bottom edge, is recovered more clearly, transitioning from an artifact-dominated outline to a well-resolved boundary. Overall, this comparison shows that suppressing reflection-induced (deterministic) phase distortion is crucial for recovering genuine object boundaries in coherent THz Fourier imaging, particularly in vignetted off-axis regions; in the following, we extend this analysis to a highly reflective metallic sample, where strong sample-induced reflections introduce an additional—and often dominant—source of coherent artifacts.

For the highly reflective sample, the influence of coherent reflections becomes even more pronounced. The metallic planar  Siemens star exhibits strong specular reflection at THz frequencies, which %
couples efficiently to internal reflection pathways within the imaging system. The sample was illuminated by a collimated beam and was positioned with its surface nearly perpendicular to the propagation direction of the beam, such that specularly reflected waves propagated back under a slight angle to the original beam path %
and could readily re-enter the optical system. As a consequence, sample-induced reflections and system-induced reflections became strongly coupled, forming closed interference loops that gave rise to coherent artifacts.

Without the complementary filters, these coupled reflections produce pronounced stripe-like interference and blurred features (Fig.~\ref{fig6}(d)), %
most evident near the center of the field of view where the fine metallic fingers of the Siemens star appear broadened and poorly resolved; zoomed-in views of this region (Fig.~\ref{fig6}(d),(g)) %
enable direct visual comparison with and without the filter pair.

With the filters inserted, the dominant reflection pathways are suppressed: as shown in Fig.~\ref{fig6}(g), %
stripe-like artifacts are much diminished and the metallic fingers are recovered with markedly improved sharpness and contrast.

To quantitatively assess this improvement, one-dimensional circular profiles are extracted along the pathways indicated in the zoomed views of Fig.~\ref{fig6}(d) and (g). %
Two representative feature sizes are analyzed: the conical metallic stripes of the Siemens star at positions with widths of 1~mm and 2~mm. The corresponding profile locations are marked by blue dashed lines (1~mm) and red dashed lines (2~mm) in the zoomed images. The resulting one-dimensional profiles are plotted in Fig.~\ref{fig6}(e) and (h), %
where solid curves with matching colors represent the respective stripe widths. In the filtered case, %
especially the blue curve (for the 1-mm stripe width) exhibits enhanced contrast and a more faithful representation of the true feature widths, confirming that the suppression of coherent reflections leads to a substantial improvement in spatial resolution and image fidelity. Quantitatively, the peak-to-trough modulation (central $100^\circ$--$200^\circ$ range) increases from 24.1~dB to 29.5~dB at the 1-mm linewidth ($\approx$5.4~dB gain), while the 2-mm linewidth is essentially unchanged (37.3~dB vs.\ 38.8~dB) -- the larger full-range discrepancy there (57.6~dB vs.\ 42.5~dB) reflects edge vignetting, not a true contrast difference.

These results demonstrate that frequency-selective isolation suppresses the interference pathways within the coherent imaging system, particularly for highly reflective planar samples where strong specular reflections couple efficiently to internal system reflections. The complementary filter pair reduces reflection-induced (deterministic) amplitude and phase distortion in phase-sensitive aspects of Fourier reconstruction. %

\begin{figure}
\centering
\includegraphics[width=9 cm]{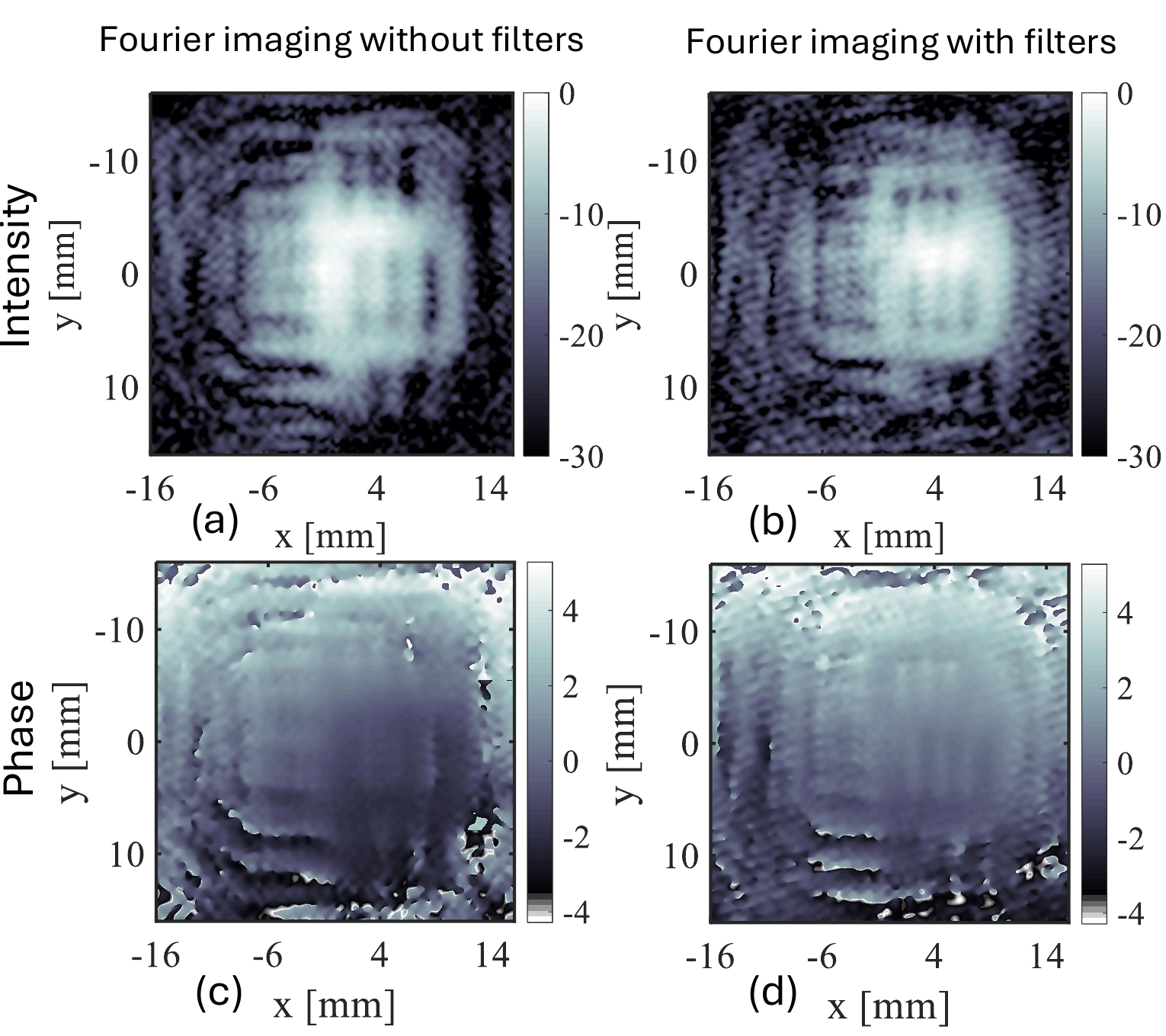}
\caption{Geometry-dependent phase behavior in coherent CW THz imaging with and without the complementary filter pair. 
(\textbf{a}, \textbf{b}) Intensity images reconstructed using Fourier imaging without and with filtering, respectively. 
(\textbf{c}, \textbf{d}) Corresponding phase maps. 
 \label{fig7}}
\end{figure}

Results of the third imaging test with another object, a planar metallic USAF test pattern, are displayed in Fig.~\ref{fig7}. While the figure shows amplitude and phase images, we concentrate in the following on the evaluation of the phase images. With regard to the amplitude images, we only want to point out the appearance of a global barrel distortion. It is also present in the images of Fig.~\ref{fig6}, but is more apparent in Fig.~\ref{fig7} for the give grey-scale presentation. The barrel distortion indicates that the focal area of lens L2 is not a plane but a curved surface. The planar raster-scanning of the detector hence does not perfectly follow the real spatial contour of the focus region which leads to the barrel distortion upon image reconstruction.

Coming to the phase images, one observes that the reconstructed phase map without filtering (Fig.~\ref{fig7}(c)) exhibits limited contrast and increased spatial nonuniformity, indicating sensitivity to residual reflections and wavefront distortions. Upon insertion of the complementary filter pair (Fig.~\ref{fig7}(d)), a systematic shift of the effective focal plane by approximately 2~mm is observed. This value is a direct experimental measurement, not a computed or inferred quantity: the detector position giving maximum signal was located separately, with and without the filters inserted, by translating the detector near the nominal focal plane of lens L2 and identifying the position of the intensity maximum; the reported 2~mm is the measured offset between these two maximum-signal positions. A continuous axial (through-focus) scan was not recorded, so only this single maximum-position estimate is available per condition, not a full depth profile, corresponding to approximately four wavelengths at 600~GHz. This shift exceeds what can be attributed to the filter alone -- the PI membrane and metamaterial phase response together account for a sub-wavelength displacement of less than 0.5~mm -- and we hypothesize is primarily related to the suppression of standing-wave interference, which in the unfiltered case can distort the detected intensity distribution and displace the apparent signal maximum from the true focal position. Filter tilt (30--45$^\circ$) and the associated beam walk-off may also perturb the reconstructed wavefront curvature and contribute to the shift; we cannot fully exclude this possibility, and independent axial through-focus scan data that would allow direct verification of the apparent-maximum-shift mechanism were not acquired. Both filters are positioned at a fixed location near the detector (within approximately 1~cm, see above), well downstream of the object itself, so filter position relative to the object plane is not a relevant degree of freedom here; a per-filter decomposition of F1's versus F2's individual contribution to the shift is not available. As qualitative, not quantitative, support that standing-wave structure in this receiver can displace an apparent intensity maximum upon filter insertion, we note that the residual interference feature in the delay-domain analysis of Fig.~\ref{figdelay}(c) shifts by approximately 9.9~mm between the unfiltered and filtered (F1+F2) traces -- not a direct prediction of the 2-mm imaging shift, since the two measurements probe different optical paths, but consistent with the same underlying mechanism of a standing wave displacing an apparent maximum-signal position.
After correction for this shift in the computational image reconstruction, the filtered Fourier image upon close inspection shows improved phase contrast -- best identified by the observation that the vertical lines of the test pattern only become visible upon filtering -- and it exhibits a more spatially uniform phase distribution -- manifested by the disappearance of the broad dark-grey stripes on the right side of the field-of-view upon filtering. This indicates that while the filters introduce a phase offset, they simultaneously suppress reflection-induced wavefront perturbations. As a result, Fourier reconstruction leads to more stable phase retrieval even under low dynamic-range conditions.

\section{Discussion and Conclusion}

In this work, we have presented a complementary metamaterial filter pair designed for frequency-selective isolation in coherent CW THz systems operating at 300~GHz and 600~GHz. The filters exhibit strong stopband suppression with low insertion loss in the passband and can be readily integrated into sub-harmonic heterodyne receiver architectures. System-level demonstrations in various imaging configurations confirm that the proposed isolation strategy effectively suppresses parasitic reflections and standing-wave-induced artifacts, resulting in improved spectral baseline stability and enhanced image fidelity.

A central finding of this study is that the impact of frequency-selective isolation extends beyond amplitude filtering. Insertion of the membrane-based metamaterial structures introduces a spatially smooth and controlled phase delay while simultaneously suppressing reflection-induced wavefront distortions. In Fourier imaging, this reduces reflection-induced (deterministic) phase distortion and enhances the validity of the reconstruction model. This is especially valuable  under limited-dynamic-range conditions. 

Beyond spectral isolation, the controlled transmission phase delay introduced by filter insertion (Appendix~B) indicates that the ultra-thin membrane-based metamaterials have broader functional potential. The combination of low insertion loss, mechanical flexibility, and controllable phase response suggests that such structures may serve as compact phase-engineering elements in terahertz systems. By tailoring the resonant geometry and spatial arrangement of the metamaterial elements, frequency-selective membranes could enable controlled wavefront shaping for applications including compressive sensing, structured illumination, and super-resolution imaging \cite{Yuan_IRMMW}, as well as spatial-light-modulator phase-shifter pixels or frequency-selective switching elements.

Future work may further optimize such metasurfaces via physics-guided inverse design, jointly targeting stopband isolation, insertion loss, polarization selectivity, and phase response -- a multi-parameter problem well suited to modern computational frameworks.

In summary, this complementary filter pair offers a practical route to reflection management in sub-harmonic coherent THz platforms, combining frequency-selective isolation with controllable phase response for next-generation THz spectroscopy and imaging systems.

\section*{Acknowledgment}
This research work was funded by DFG project RO 770/48-1. Meng Zhang was funded by the Marie
Skłodowska-Curie grant (No 765426, TeraApps). Daniel Erni acknowledges the partial support by the Deutsche Forschungsgemeinschaft (DFG) through CRC/TRR 196 MARIE (287022738) under project M03.

The authors would like to thank Uwe Lischka for the fabrication of the filters in the cleanroom. We also acknowledge Alvydas Lisauskas and Jakob Holstein for providing the TeraFET detector with the sub- and superstrate lenses. The authors further thank Aparajita Bandyopadhyay for insightful discussions during the preparation of the manuscript.

The data that support the findings of this study are available from the corresponding author upon reasonable request.

\appendices
\section{Quasi-optical measurement geometry and standing-wave beam paths identified in Fig.~\ref{fig4}}
\label{appendixa}
\begin{figure}
\centering
\includegraphics[width=9 cm]{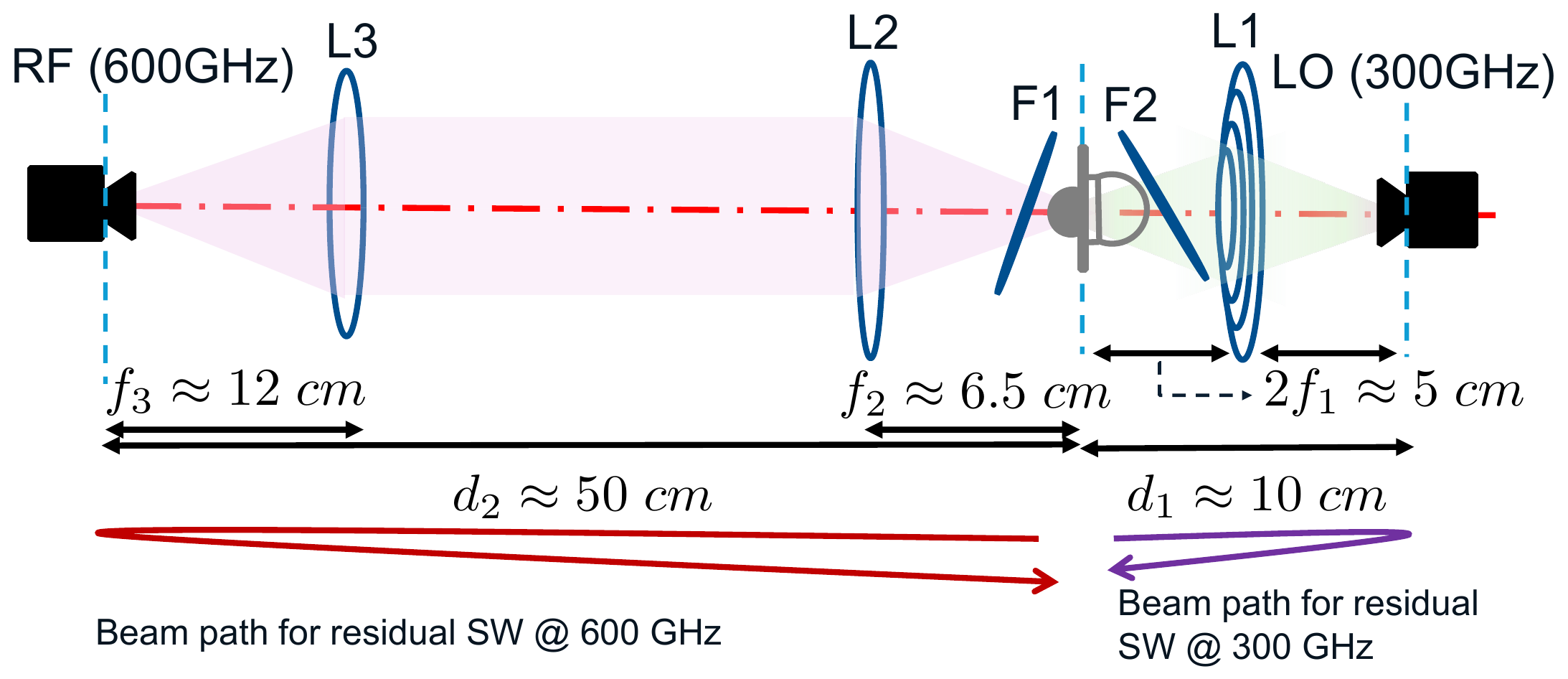}\\[2mm]
\includegraphics[width=7 cm]{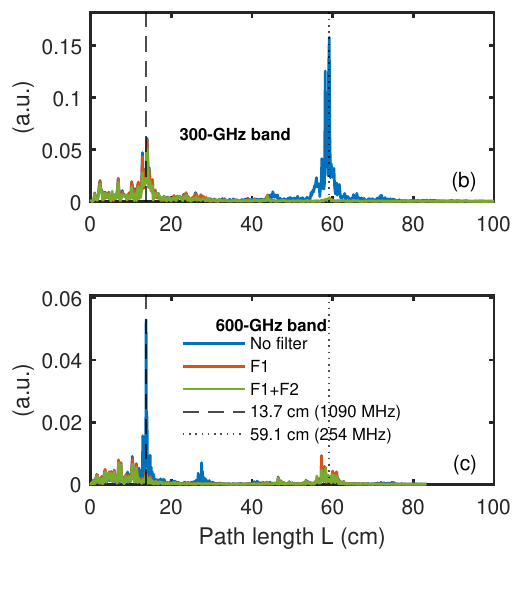}
\caption{(\textbf{a}) Quasi-optical receiver geometry: the RF beam (600~GHz) couples to the detector via lenses L2--L3 and a Si substrate lens (front side); the LO beam (300~GHz) via Fresnel lens L1 and a PTFE superstrate lens (rear side). Filters F1, F2 sit immediately in front of/behind the detector. The colored solid lines denote significant reflection pathways, including the lower labeled arrows indicating the residual self-reflection loops in the RF arm ($d_2\approx50$~cm) and LO arm ($d_1\approx10$~cm) quantified in Table~\ref{tab:ripple}. (\textbf{b}, \textbf{c}) Delay-domain analysis of the raw spectra underlying Fig.~\ref{fig4}, validating the path-length assignment: in each band, the dominant peak ($L\approx59.1$~cm/254~MHz in \textbf{b}; $L\approx13.7$~cm/1090~MHz in \textbf{c}) is removed by the filter in the opposite arm (F1, F2 respectively), while the weaker, residual same-arm peak persists essentially unchanged with both filters inserted.
\label{figa1}\label{figdelay}}
\end{figure}

Fig.~\ref{figa1} illustrates the quasi-optical sub-harmonic heterodyne measurement configuration used for the characterization of the complementary frequency-selective filters in Sec.~\ref{3.2} and the imaging experiments. The RF signal at 600~GHz propagates through the front-side optical path and is focused onto the detector using the lens system L3--L2 and a 4-mm-diameter hyper-hemispherical Si substrate lens. The total RF propagation distance between source and detector is approximately 50~cm. The LO signal at 300~GHz illuminates the detector from the opposite side through an independent 4$f_1$ optical path (composed of Fresnel lens L1 with a focal length of 2.5~cm and an 8-mm-diameter hemispherical superstrate PTFE lens) with a total propagation distance of approximately 10~cm. The complementary filters F1 and F2 are positioned immediately in front of resp. behind the detector: F1 is designed to reject radiation near 300~GHz while transmitting the RF frequency, whereas F2 rejects radiation near 600~GHz while remaining highly transmissive at the LO frequency of 300~GHz.

For radiation propagating along the primary propagation paths, reflections occur at the detector, filters, dielectric lenses, and source modules. Representative (and -- as one finds from the analysis of Fig.~\ref{fig4}, see below -- dominant) reflection paths are indicated by the trajectories indicated in color in the lower part of Fig.~\ref{figa1}. These reflections can generate standing-wave interference patterns which reveal themselves in the frequency response measured by the detector.

For a standing-wave cavity with effective round-trip propagation length $2L$, the corresponding frequency spacing between two subsequent interference maxima is given by
\begin{equation}
\label{standinbg}
\Delta f = \frac{c}{2L},
\end{equation}
where $c$ denotes the speed of light in free space.

In each of the spectra of Fig.~\ref{fig4}(a) and (b), one identifies by inspection two dominant modulation frequencies, approximately 1.1~GHz and 256~MHz. A precise determination via the delay-domain (Fourier-transform) analysis of the raw spectra (Fig.~\ref{figdelay}(b,c)) gives $\Delta f = 1090$~MHz and $254$~MHz, respectively. Using relation (\ref{standinbg}), a standing-wave period of 1090~MHz corresponds to an effective one-way propagation length of $L = c/(2\Delta f)\approx 13.7~\mathrm{cm}$, while a standing-wave period of 254~MHz corresponds to $L \approx 59.1~\mathrm{cm}$. These are optical path lengths and are therefore somewhat larger than the purely geometric distances quoted above (10~cm for the LO arm, 50~cm for the RF arm); the difference is accounted for by the extra optical length introduced by the dielectric lenses (the PTFE superstrate lens and the Si substrate lens) traversed in each arm. On this basis, the 13.7-cm feature is identified with round trips confined to the short, $\approx$10-cm LO arm, and the 59.1-cm feature with round trips confined to the long, $\approx$50-cm RF arm.

The dominant mechanism is cross-leakage between the two arms: because the detector does not fully absorb either beam, unabsorbed radiation continues to propagate into the \emph{opposite} arm, where it is reflected and returns through the detector to interfere with the directly detected signal, so that the round-trip length is set by the arm into which the radiation has leaked rather than by the arm of the originating source.

The effect of the filters is then as follows. In Fig.~\ref{fig4}(a), one observes strong suppression of the dominant 254-MHz modulation by F1 which (at least partially) blocks the LO radiation in the 300-GHz band. The additional insertion of F2 does not enhance the suppression of the 254-MHz modulation and also does not suppress the 1090-MHz modulation because it is transparent to the LO radiation.

In Fig.~\ref{fig4}(b), the dominant 1090-MHz modulation is strongly suppressed by F2 which (at least partially) blocks the RF radiation in the 600-GHz band. Here, the additional insertion of F1 does not suppress any of the two modulations because F1 lets the RF signal pass.

This assignment is independently confirmed by a delay-domain (Fourier-transform) analysis of the raw power-detection spectra underlying Fig.~\ref{fig4}, shown in Fig.~\ref{figdelay}: inserting the filter located in the arm into which the radiation leaks removes the corresponding peak, while the weaker same-arm feature remains unaffected.

\section{Simulated transmission of F1 and F2 at the operational tilt angle}
\label{appendixb}

\begin{figure}
\centering
\includegraphics[width=6 cm]{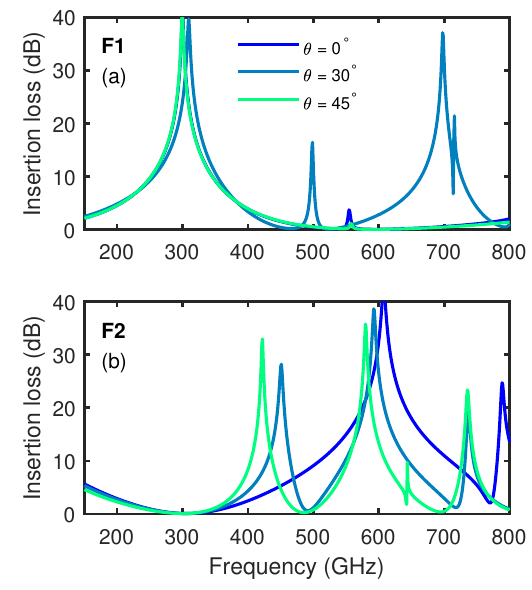}
\caption{Simulated insertion loss $L_T=-10\log_{10}(T)$ versus frequency at the azimuthal orientation $\phi=0^\circ$, for polar tilt angles $\theta=0^\circ$ (normal incidence), $30^\circ$, and $45^\circ$, spanning the operational tilt range used in Sec.~\ref{3.2}. (\textbf{a}) Filter F1. (\textbf{b}) Filter F2.}
\label{figtilt}
\end{figure}

The filters are operated tilted by approximately 30--45$^\circ$ from the optical axis (Sec.~\ref{3.2}), whereas the characterization in Figs.~\ref{fig2} and~\ref{fig3} was performed at normal incidence. To address this, we performed full-wave simulations of F1 and F2 at oblique incidence, for the azimuthal orientation $\phi=0^\circ$ (matching the $0^\circ$-polarization reference used in Fig.~\ref{fig2}) and polar tilt angles $\theta=0^\circ$ (normal incidence), $30^\circ$, and $45^\circ$. Fig.~\ref{figtilt} shows the resulting insertion-loss spectra.

For F1, the stopband near 300~GHz shows only a modest shift with increasing tilt, and the passband near 600~GHz remains essentially unaffected over the full $0$--$45^\circ$ range. Additional narrow resonances appear near 490--720~GHz at $\theta=30^\circ$ that are not present at $\theta=0^\circ$ or $45^\circ$; these do not coincide with the 600-GHz passband.

For F2, the stopband near 600~GHz is essentially unshifted between $\theta=0^\circ$ and $30^\circ$, but shifts substantially -- by approximately 180~GHz, down to $\approx$420~GHz -- at $\theta=45^\circ$.

In the realistic receiver, the tilt angle of each filter within the stated 30--45$^\circ$ range is set manually as a trade-off between maximizing the passband transmission and keeping the reflected beam clear of the optical path (Sec.~\ref{3.2}), rather than fixed a priori. This calibration procedure is consistent with the angle dependence shown here: for F2 in particular, operating toward the lower end of the range (closer to $30^\circ$) preserves the design stopband near 600~GHz, whereas the stopband isolation at the intended frequency is reduced if the filter is tilted toward $45^\circ$.

\section{Comparison with prior THz frequency-selective filters}
\label{appendixc}

Table~\ref{tab:comparison} compares the present filters with representative prior THz FSS/metal-mesh filters.

\begin{table}[t]
\centering
{\renewcommand{\arraystretch}{0.85}\setlength{\tabcolsep}{3pt}

\caption{Comparison with representative prior THz frequency-selective filters (n/r: not reported in the cited source).}
\label{tab:comparison}
\begin{tabular}{p{0.9cm} p{2.1cm} p{1.2cm} p{1.1cm} p{2.3cm}}
\hline\hline
Ref. & Substrate & Freq. & Passband IL & System-level application\\
\hline
This work & PI, 5~$\mu$m & 300 \& 600~GHz & $<$1~dB & Heterodyne receiver \\
\cite{Ade2006metalmesh} & Metal mesh (review) & FIR/sub-mm & n/r & Astronomy instr. \\
\cite{9706347} & Multilayer FSS & THz band & $\approx$1~dB & Standalone filter \\
\cite{WU2023115563} & PDMS & 360--446~GHz & $<$0.3~dB & Standalone filter \\
\hline\hline
\end{tabular}}
\end{table}

\providecommand{\noopsort}[1]{}\providecommand{\singleletter}[1]{#1}%

\end{document}